\newcommand{\bea}{\begin{eqnarray}}
\newcommand{\eea}{\end{eqnarray}}
\newcommand{\nn}{\nonumber \\}

\documentclass[%
 reprint,
 amsmath,amssymb,
 aps,
 prd,
 showkeys
]{revtex4-1}

\usepackage{float}
\usepackage[USenglish]{babel}
\usepackage{graphicx}
\usepackage{dcolumn}
\usepackage{bm}
\usepackage{hyperref}

\begin{document}

\preprint{APS/123-QED}

\title{Superstatistics and the effective QCD phase diagram}

\author{Alejandro Ayala$^{1,2}$, Martin Hentschinski$^3$ L. A. Hern\'andez$^{1,2}$, M. Loewe$^{4,2,5}$, and R. Zamora$^{6,7}$}
\affiliation{%
$^1$Instituto de Ciencias Nucleares, Universidad Nacional Aut\'onoma de M\'exico, Apartado Postal 70-543, CdMx 04510, Mexico.\\
$^2$Centre for Theoretical and Mathematical Physics, and Department of Physics, University of Cape Town, Rondebosch 7700, South Africa.\\
$^3$Departamento de Actuaria, F\'isica y Matem\'aticas, Universidad de las Am\'ericas Puebla,
Santa Catarina M\'artir, San Andr\'es Cholula, 72820 Puebla, Mexico.\\
$^4$Instituto de F\'isica, Pontificia Universidad Cat\'olica de Chile, Casilla 306, Santiago 22, Chile.\\
$^5$Centro Cient\'ifico-Tecnol\'ogico de Valpara\'iso CCTVAL, Universidad T\'ecnica Federico Santa Mar\'ia, Casilla 110-V, Valapara\'iso, Chile\\
$^6$Instituto de Ciencias B\'asicas, Universidad Diego Portales, Casilla 298-V, Santiago, Chile.\\
$^7$Centro de Investigaci\'on y Desarrollo en Ciencias Aeroespaciales (CIDCA), Fuerza A\'erea de Chile, Casilla 8020744, Santiago, Chile.
}%


\begin{abstract}
We study the effect of a  partially thermalized scenario for  chiral symmetry restoration at finite temperature and quark chemical potential, and in particular for the position of the critical end point in an effective description of the QCD phase diagram. We show that these effects cause the critical end point to be displaced towards larger values of temperature and lower values of the quark chemical potential as compared to the case when the system can be regarded as completely thermalized. We conclude that these effects may be important for relativistic heavy ion collisions where the number of subsystems making up the whole interaction volume can be linked to the finite number of participants in the reaction.
\end{abstract}

\pacs{Valid PACS appear here}
\maketitle


\section{Introduction}\label{I}

The usual thermal description of a relativistic heavy-ion collision relies on the assumption that the produced matter reaches equilibrium
after some time from the beginning of the reaction. This equilibrium is characterized by values of temperature $T$ and baryon chemical potential $\mu$ which are taken as common within the whole interaction volume. The  system's evolution is subsequently described by the time evolution of the temperature down to kinetic freeze-out, where particle spectra are established. This picture rests on two ingredients: the validity of Gibbs-Boltzmann statistics and a system's adiabatic evolution.

Although, for expansion rates not too large compared to the interaction rate, the adiabatic evolution can perhaps be safely assumed, it is well known that the Gibbs-Boltzmann statistics can be applied only to systems in the thermodynamical limit, namely, long after the relaxation time has elapsed and randomization has been achieved within the system's volume. In the case of a relativistic heavy-ion collision, the reaction starts off from nucleon-nucleon interactions. This means that the entire reaction volume is made, at the beginning, of a superposition of interacting nucleons pairs. If thermalization is achieved, it seems natural to assume that this starts off in each of the interacting nucleon pair subsystems, and later spread to the entire volume. In this scenario, the temperature and chemical potential within each subsystem may not be the same for other subsystems and thus a superposition of statistics, one in the usual Gibbs-Boltzmann sense for particles in each subsystem, and another one, for the probability to find particular values for $T$ and $\mu$ for different subsystem, seems appropriate. This is described by the so-called superstatistics scenario. 

An important characteristic of superstatistics is that a nonextensive behavior naturally arises due to fluctuations in $T$ or $\mu$ over the system's volume. This feature could be of particular relevance when studying the position of the critical end point (CEP) in the QCD phase diagram, where one resorts to measuring ratios of fluctuations in conserved charges, with the expectation that the volume factor cancels out in the ratio. If thermalization is not complete, this expectation cannot hold and a more sophisticated treatment is called for. 

From the theoretical side, efforts to locate the CEP employing several techniques such as finite energy sum rules~\cite{QCDsumrules}, Schwinger-Dyson equations, functional renormalization methods, holography, and effective models~\cite{values,Scoccola,Ayala-Dominguez,Xin,Fischer,Lu,LSModel,Ayala2,Ayala5,Shi,Contrera,NJL,Cui,Datta,Carlomagno,Knaute,Antoniou,Rougemont,RMF} where recently carried out. In all of these cases, a full thermalization over the whole reaction volume has been assumed. From the experimental side, the STAR BES-I program has recently studied heavy-ion collisions in the energy range 200 GeV $ > \sqrt{s_{NN}} > $ 7.7 GeV~\cite{BESI}. Future experiments~\cite{BESII,FAIR,NICA} will continue to thoroughly explore the QCD phase diagram, using different system sizes and varying the temperature and baryon density using different collision energies down to about $\sqrt{s_{NN}}\simeq 5$ GeV.

The superstatistics scenario has been explored in the context of relativistic heavy-ion collisions in a wide range of papers, {\it e.g.} Refs.~\cite{Wilk2,Wilk3,Wilk4,Rybczynski:2014cha,
Wong:2015mba,Wilk5,Bialas:2015pla,Bhattacharyya:2015hya,Bialas:2015oua,Rozynek:2016ykp,Tripathy:2016hlg,Grigoryan:2017gcg,Bhattacharyya:2017hdc,Khuntia:2017ite,Tripathy:2017nmo,Ishihara:2017txj,Wilk:2018kvg} and references therein, with a particular focus on the study of the imprints of superstatistics on particle production, using a particular version, the so-called Tsallis statistics~\cite{Tsallis}. Its use in the context of the computation of the rapidity distribution profile for stopping in heavy ion collisions has been recently questioned in Ref.~\cite{Wolschin}. It has also been implemented to study generalized entropies and a generalized Newton's law in Refs.~\cite{Obregon1, Obregon2, Obregon3}. In this work we explore the implications of superstatistics for the location of the CEP in the QCD phase diagram. The work is organized as follows. In Sec.~\ref{II} we briefly review the superstatistics scenario for the case when the fluctuating parameter is the inverse temperature. In Sec.~\ref{III} we apply a superstatistics analysis to the theory of a self-interacting boson with spontaneously broken $Z_2$ symmetry and show how the critical temperature for symmetry restoration decreases as the number of subsystems making up the whole system also decreases. In Sec.~\ref{IV} we analyze the superstatistcis scenario within the Linear Sigma Model with quarks (LSMq). We compute the corrections to the effective potential at finite temperature and baryon chemical potential and show how these produce a displacement of the CEP towards larger values of the critical temperature and lower values of the quark chemical potential, as compared to the case where a full-volume thermalization is assumed. Finally we summarize and give an outlook of the analysis in Sec.~\ref{concl}.

\section{Superstatitics}\label{II}

The superstatistics concept was nicely described in Refs.~\cite{Beck2004-I,Beck2004}. For completeness, we reproduce here the main ideas. 

For a system that has not yet reached a full equilibrium and contains space-time fluctuations of an intensive parameter $\beta$, such as the inverse temperature or the chemical potential, one can still think of dividing the full volume into spatial subsystems, where $\beta$ is approximately constant. Within each subsystem, one can apply the ordinary Gibbs-Boltzmann statistics, namely, one can use the ordinary matrix density giving rise to the Boltzmann factor $e^{-\beta \hat{H}}$, where $\hat{H}$ corresponds to the Hamiltonian for the states in each subsystem. The whole system can thus be described in terms of a space-time average over the different values that $\beta$ could take for the different subsystems. In this way, one obtains a superposition of two statistics, one referring to the Boltzmann factor $e^{-\beta \hat{H}}$ and the other for $\beta$, hence the name superstatistics.

To implement the scenario, one defines an averaged Boltzmann factor 
\bea
   B(\hat{H})=\int_0^\infty f(\beta)e^{-\beta\hat{H}}d\beta,
\label{averageBoltzmann}
\eea
where $f(\beta)$ is the probability distribution of $\beta$. The partition function then becomes
\bea
   Z&=&{\mbox{Tr}}[B(\hat{H})]\nonumber\\
    &=&\int_0^\infty B(E)dE,
\label{partitionfuntion}
\eea
where the last equality holds for a suitably chosen set of eigenstates of the Hamiltonian. 

When the subsystems can all be described with the same  probability distribution~\cite{Obregon1}, a possible choice to distribute the random variable $\beta$ is the $\chi^2$  distribution
\bea
   f(\beta)=\frac{1}{\Gamma(N/2)}\left(\frac{N}{2\beta_0}\right)^{N/2}\beta^{N/2-1}e^{-N\beta /2\beta_0},
\label{chisquared}
\eea
where $\Gamma$ is the Gamma function, $N$ represents the number of subsystems that make up the whole system and 
\bea
   \beta_0\equiv\int_0^\infty\beta f(\beta)d\beta = \langle\beta\rangle,
\label{betazero}
\eea
is the average of the distribution. The $\chi^2$ is the distribution that emerges for a random variable that is made up of the sum of the squares of random variables $X_i$, each of which is distributed with a Gaussian probability distribution with vanishing average and unit variance. This means that if we take
\bea
   \beta = \sum_{i=1}^N X_i^2,
\label{sumrandomX}
\eea
then $\beta$ is distributed according to Eq.~(\ref{chisquared}). Moreover, its variance is given by
\bea
   \langle\beta^2\rangle -\beta_0^2 = \frac{2}{N}\beta_0^2.
\label{variance}
\eea
Given that $\beta$ is a positive-definite quantity, thinking of it as being the sum of positive-definite random variables is an adequate model. Notice however that these variables do not necessarily correspond to the inverse temperature in each of the subsystems. However, since the use of the $\chi^2$ distribution allows for an analytical treatment, we hereby take this as the distribution to model the possible values of $\beta$.

To add superstatistics effects to the dynamics of a given system, we first find the effective Boltzmann factor. This is achieved by taking Eq.~(\ref{chisquared}) and substituting it into Eq.~(\ref{averageBoltzmann}). Integration over $\beta$, leads to
\begin{equation}
 B(\hat{H})=(1+\frac{2}{N}\beta_0 \hat{H})^{-\frac{N}{2}}.
 \label{generalBoltzmanfactor}
\end{equation}
Notice that in the limit when $N\to\infty$, Eq.~(\ref{generalBoltzmanfactor}) becomes the ordinary Boltzman factor. For large but finite $N$, Eq.~(\ref{generalBoltzmanfactor}) can be expanded as
\begin{eqnarray}
 B(\hat{H})&=&\Big [1+\frac{1}{2}\Big(\frac{2}{N}\Big)\beta_0^2\hat{H}^2-\frac{1}{3}\Big(\frac{2}{N}\Big)^2\beta_0^3\hat{H}^3+ \cdots \Big]\nonumber\\
&\times&e^{-\beta_0 \hat{H}}
\label{Boltzmanfactorapprox}
\end{eqnarray}
Working up to first order in $1/N$, Eq.~(\ref{Boltzmanfactorapprox}), can be written as~\cite{Beck2004-I}
\begin{eqnarray}
   B(\hat{H})=
\left[1+ \frac{\beta_0^2}{N} \left(\frac{\partial}{\partial \beta_0}\right)^2
\right]e^{-\beta_0 \hat{H}}.
\end{eqnarray}
Therefore, the partition function to first order in $1/N$ is
given by
\begin{equation}
Z=\left[1+ \frac{\beta_0^2}{N} \left(\frac{\partial}{\partial \beta_0}\right)^2
\right] Z_0  
\end{equation}
with
\begin{eqnarray}
Z_0=
e^{-\texttt{V}\beta_0 V^{\text{eff}}},
\label{partitionfunctionnormal}
\end{eqnarray}
where $\texttt{V}$ and $V^{\text{eff}}$ are the system's volume and effective potential, respectively. After a bit of straightforward algebra we write the expression for the partition function  in terms of $T_0=1/\beta_0$ as
\begin{align}
 Z&=\left[1+ \frac{\beta_0^2}{N} \left(\frac{\partial}{\partial \beta_0}\right)^2
 \right] Z_0 
 \nonumber \\
  &=Z_0\left[1+\frac{2T_0}{N Z_0} \left( \frac{\partial Z_0 }{\partial T_0} + \frac{T_0}{2} \frac{\partial^2 Z_0 }{\partial T_0^2} \right)\right],
 \label{partfuncsuper}
\end{align}
and therefore
\begin{eqnarray}
\ln [Z]&=&\ln[Z_0]\nonumber\\
&+&\ln\Bigg[1+\frac{2T_0}{N Z_0} \left( \frac{\partial Z_0}{\partial T_0} + \frac{T_0}{2} \frac{\partial^2 Z_0}{\partial T_0^2} \right)\Bigg].
\label{logZsuper}
\end{eqnarray}
The question we set out to answer is how the CEP position changes when considering corrections coming from the second term in Eq.~(\ref{logZsuper}). In the realm of effective QCD models, the answer should be provided within a theory that involves meson as well as quarks degrees of freedom (d.o.f). However, before we delve into this problem, it is convenient to make a first exploratory study within a simpler theory involving only one boson d.o.f., for which the $Z_2$ symmetry is spontaneously broken at $T_0=0$ but restored at high temperature. As we proceed to show, the correction term coming from superstatistics produces a modification of the temperature for symmetry restoration.

\section{Superstatistics in the $\phi^4$ theory}\label{III}

To study the effects of superstatistics on the symmetry restoration temperature, we first consider a theory of a self-interacting scalar field that undergoes spontaneous symmetry breaking. This model is described by the Lagrangian
\bea
   {\mathcal{L}}=\frac{1}{2}(\partial^{\mu}\phi)(\partial_{\mu}\phi)
   +\frac{a^{2}}{2}\phi^2-\frac{\lambda}{4!}
   \phi^{4},
\label{lagrangian}
\eea
where $a^2>0$ is the squared mass parameter and $\phi$ is a real, self-interacting scalar field, with an interaction strength $\lambda>0$. To allow for a spontaneous breaking of symmetry, we let the $\phi$ field to develop a vacuum expectation value $v$
\bea
   \phi \rightarrow \sigma + v,
\label{shift}
\eea
which can later be taken as the order parameter of the theory. After this shift, the Lagrangian can be rewritten as
\bea
   {\mathcal{L}} &=& \frac{1}{2}(\partial^\mu\sigma)(\partial_{\mu}\sigma)-\frac{1}
   {2}\left(\lambda v^{2}-2a^{2} \right)\sigma^{2} - \frac{\lambda}{4!}\sigma^{4}\nn 
   &+&\frac{a^{2}}{2}v^{2}-\frac{\lambda}{4!}v^{4}.
  \label{higgsl}
\eea
From Eq.~(\ref{higgsl}) we see that the mass of the $\sigma$ field is given by
\begin{equation}
  m^2_{\sigma}=\lambda v^{2}-2a^{2}.
\label{masses}
\end{equation}
In this work we consider the effective potential beyond the mean-field approximation. We include radiative corrections up to the ring diagrams contribution. All matter terms are computed in the high-temperature approximation. The effective potential is given by \cite{Ayala-Sahu}
\begin{align}
    V^{\text{eff}}(v,T_0)&=-\frac{(a^2+\delta a^2)}{2}v^2
    +\frac{(\lambda+\delta \lambda)}{4!}v^4\nonumber \\
    &-\frac{m_\sigma^4}{64\pi^2}\Big[ \ln \Big( \frac{ a^2}{4\pi T_0^2}\Big)-\gamma_E+\frac{1}{2} \Big]\nonumber \\
    &-\frac{\pi^2 T_0^4}{90}+\frac{m_\sigma^2 T_0^2}{24}\nonumber \\
    &-\frac{[m_\sigma^2+\Pi(T_0)]^{3/2} T_0}{12\pi},
     \label{finalhigs}
\end{align}
where we have chosen the renormalization scale $\mu=ae^{-1/2}$, $\gamma_E$ is the Euler-Mascheroni constant and we have introduced the leading temperature plasma screening effects for the boson's mass squared, encoded in the boson's self-energy
\bea
   \Pi=\lambda \frac{T_0^2}{24}.
\label{self}
\eea
Equation~(\ref{finalhigs}) contains the counter-terms $\delta a^2$ and $\delta\lambda$, given by
\begin{equation}
 \delta a^2= -a^2 \lambda \left[\frac{7+2\ln(2)}{64\pi^2}\right],
\end{equation} 
and 
\begin{equation}
 \delta \lambda= -3\lambda^2\left[\frac{3+2\ln(2)}{64\pi^2}\right],
\end{equation}
which ensure that the vacuum piece of the one-loop radiative corrections do not shift the minimum or the $\sigma$-mass from their tree-level values~\cite{Carrington}. To implement superstatistics in the analysis, we substitute Eq.~(\ref{finalhigs}) into Eq.~(\ref{partitionfunctionnormal}).   In particular with
\begin{equation}
\label{eq:Z0eff}
Z_0=e^{-\frac{\texttt{V}}{T_0} V^{\text{eff}}(v,T_0)},
\end{equation}
we define an effective potential which takes into account effects due to superstatistics through,
\begin{align}
Z & = e^{-\frac{\texttt{V}}{T_0} V^{\text{eff}}_{\text{sup}}(v,T_0, N)}. 
\end{align}
which yields
\begin{eqnarray}
\label{eq:lnZsuper}
  V^{\text{eff}}_{\text{sup}}(v,T_0, N)&=&-\frac{T_0}{\texttt{V}}\ln [Z]\nonumber\\
  &=& V^{\text{eff}}(v,T_0) \nonumber\\
  &-&\frac{T_0}{\texttt{V}}\ln\Bigg[1+\frac{2T_0}{N Z_0} \left( \frac{\partial Z_0}{\partial T_0} + \frac{T_0}{2} \frac{\partial^2 Z_0}{\partial T_0^2} \right)\Bigg].\nonumber\\
\end{eqnarray}
with $Z_0$ given by Eq.~\eqref{eq:Z0eff}. 
Depending on the chosen values for the parameters $a$ and $\lambda$, the effective potential $V_{\text{sup}}^{\text{eff}}$ in Eq.~(\ref{eq:lnZsuper}) exhibits both first- and second-order phase transitions.  For a first-order phase transition, $V_{\text{sup}}^{\text{eff}}$ has two degenerate minima at the critical temperature $T_c$: one located at $v=0$ and the other one at a finite value of $v$. In contrast, for a second-order transition, $V_{\text{sup}}^{\text{eff}}$ has a single minimum and vanishing curvature at $v=0$, at the critical temperature $T_c$. This is illustrated in Fig.~1.

Figure~2 shows the change of $T_c$, referred to as the critical temperature for a completely thermalized system $T_c^0$, as a function of the number of subsystems $N$ making up the whole system. The dots correspond to the case of a first-order phase transition, computed with $\lambda=5$ and $a=50$~MeV. The squares correspond to a second-order phase transition, computed with $\lambda=0.01$ and $a=50$~MeV. Notice that in both cases $T_c/T_c^0<1$, with the ratio decreasing with decreasing $N$. For increasing $N$, $T_c/T_c^0$ approaches unity, both in the case of first- and second-order phase transitions. The limit $T_c/T_{c}^0=1$ requires considerably larger values of $N$ for the latter. This behavior is to be expected since for second-order phase transitions, in the thermodynamical limit, large spatial correlations appear at the critical temperature. One may therefore anticipate that the critical temperature attains the value corresponding to a fully thermalized system only when $N$ becomes infinity.
\begin{figure}[t!]
 \includegraphics[scale=0.6]{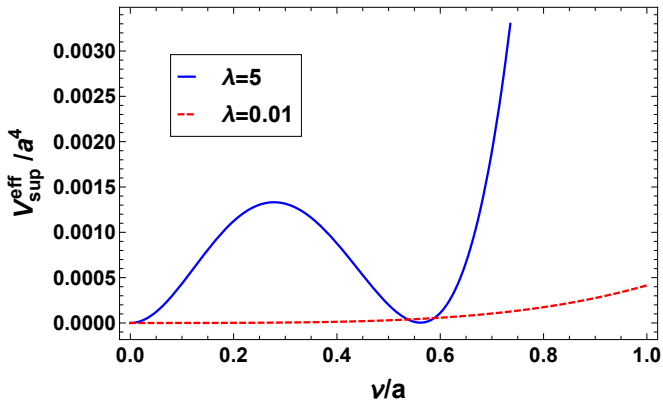}
 \label{fig1}
 \caption{Effective potential as a function of the vacuum expectation value $v$ for $N=100$ and $a=50$ MeV. The solid line corresponds to a first order transition and the dashed line to a second order transition, at their corresponding critical temperatures.}
\end{figure}
\section{Superstatistics and the linear sigma model with quarks}\label{IV}

To explore the QCD phase diagram from the point of view of chiral symmetry restoration, we use an effective model that accounts for the physics of spontaneous symmetry breaking at finite temperature and density: the linear sigma model. In order to account for the fermion d.o.f. around the phase transition, we also include quarks in this model and work with the LSMq. The Lagrangian for the case when only the two lightest quark flavors are included is given by
\begin{eqnarray}
   \mathcal{L}&=&\frac{1}{2}(\partial_\mu \sigma)^2  + \frac{1}{2}(\partial_\mu \vec{\pi})^2 + \frac{a^2}{2} (\sigma^2 + \vec{\pi}^2) - \frac{\lambda}{4} (\sigma^2 + \vec{\pi}^2)^2 \nonumber \\
   &+& i \bar{\psi} \gamma^\mu \partial_\mu \psi -g\bar{\psi} (\sigma + i \gamma_5 \vec{\tau} \cdot \vec{\pi} )\psi ,
\label{lagrangian}
\end{eqnarray}
where $\psi$ is an SU(2) isospin doublet, $\vec{\pi}=(\pi_1, \pi_2, \pi_3 )$ is an isospin triplet and $\sigma$ is an isospin singlet. $\lambda$ is the boson's self-coupling and $g$ is the fermion-boson coupling. $a^2>0$ is the squared mass parameter. 

\begin{figure}[t!]
 \includegraphics[scale=0.5]{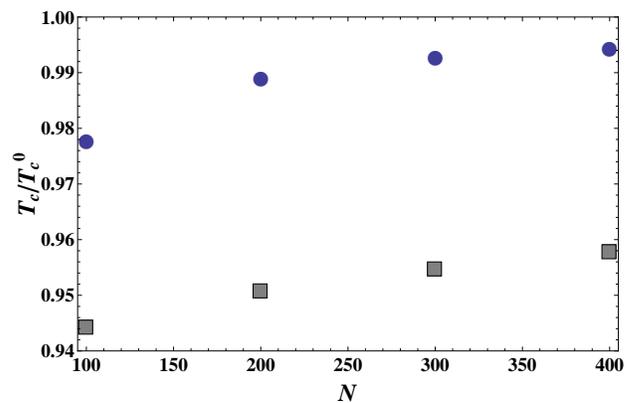}
 \label{fig2}
 \caption{Critical temperature $T_c$ for symmetry restoration, divided by the critical temperature for a completely thermalized system, $T_c^0$, as a function of the number of subsystems $N$ that make up the whole volume. The full circles correspond to the case of a first order phase transition and are computed with $\lambda=5$ and $a=50$~MeV. The squares correspond to a second order phase transition, computed with $\lambda=0.01$ and $a=50$~MeV.}
\end{figure}
To allow for an spontaneous symmetry breaking, we let the $\sigma$ field develop a vacuum expectation value $v$
\begin{equation}
   \sigma \rightarrow \sigma + v,
\label{shift}
\end{equation}
that serves as the order parameter to identify the phase transitions. After this shift, the Lagrangian can be rewritten as
\begin{eqnarray}
   {\mathcal{L}} &=&\frac{1}{2}(\partial_\mu \sigma)^2-\frac{1}
   {2}\left(3\lambda v^{2}-a^{2} \right)\sigma^{2}\nonumber \\
   &+&\frac{1}{2}(\partial_\mu \vec{\pi})^2-\frac{1}{2}\left(\lambda v^{2}- a^2 \right)\vec{\pi}^{2}+\frac{a^{2}}{2}v^{2}\nonumber \\
  &-&\frac{\lambda}{4}v^{4} + i \bar{\psi} \gamma^\mu \partial_\mu \psi 
  -gv \bar{\psi}\psi + {\mathcal{L}}_{I}^b + {\mathcal{L}}_{I}^f,
  \label{lagranreal}
\end{eqnarray}
where the sigma, the three pions and the quarks have masses given by
\begin{eqnarray}
  m^{2}_{\sigma}&=&3  \lambda v^{2}-a^{2},\nonumber \\
  m^{2}_{\pi}&=&\lambda v^{2}-a^{2}, \nonumber \\
  m_{f}&=& gv,
\label{masses}
\end{eqnarray}
respectively, and ${\mathcal{L}}_{I}^b$ and  ${\mathcal{L}}_{I}^f$ are given by
\begin{eqnarray}
  {\mathcal{L}}_{I}^b&=&-\frac{\lambda}{4}(\sigma^2 + \vec{\pi}^2)^2\nonumber \\ 
  {\mathcal{L}}_{I}^f&=&-g\bar{\psi} (\sigma + i \gamma_5 \vec{\tau} \cdot \vec{\pi} )\psi.
  \label{lagranint}
\end{eqnarray}
Equation~(\ref{lagranint}) describes the interactions among the $\sigma$, $\vec{\pi}$ and $\psi$ fields after symmetry breaking.

In order to analyze chiral symmetry restoration, we compute the finite temperature and density effective potential. In order to account for plasma screening effects, we also work up to the ring diagrams contribution. All matter terms are computed in the high temperature approximation. The effective potential is given by~\cite{RMF}
\begin{align}
    V^{\text{eff}}(v,T_0,\mu_q)&=-\frac{(a^2+\delta a^2)}{2}v^2
   +\frac{(\lambda+\delta \lambda)}{4}v^4\nonumber \\
    &  \hspace{-1.5cm} +\sum_{b=\sigma,\bar{\pi}}\Big\{-\frac{m_b^4}{64\pi^2}\Big[ \ln \Big( \frac{ a^2}{4\pi T_0^2}\Big)-\gamma_E+\frac{1}{2} \Big]\nonumber \\
    &   \hspace{-1cm}-\frac{\pi^2 T_0^4}{90}+\frac{m_b^2 T_0^2}{24}
    -\frac{(m_b^2+\Pi(T_0,\mu_q))^{3/2} T_0}{12\pi}\Big\}\nonumber \\
    & \hspace{-1.5cm} +\sum_{f=u,d}\Big\{\frac{m_f^4}{16\pi^2}\Big[ \ln \Big( \frac{ a^2}{4\pi T_0^2}\Big)-\gamma_E+\frac{1}{2}\nonumber \\
    & \hspace{-1cm} -\psi^0\Big( \frac{1}{2}+\frac{\text{i}\mu_q}{2\pi T_0} \Big)-\psi^0\Big( \frac{1}{2}-\frac{\text{i}\mu_q}{2\pi T_0} \Big)\Big]\nonumber \\
    &  \hspace{-1cm} -8m_f^2T_0^2\Big[ \text{Li}_2(-e^{\mu_q/T_0})+\text{Li}_2(-e^{-\mu_q/T_0}) \Big]\nonumber \\
    &  \hspace{-1cm} +32T_0^4\Big[ \text{Li}_4(-e^{\mu_q/T_0})+\text{Li}_4(-e^{-\mu_q/T_0}) \Big]\Big\},
    \label{finalHTpotential}
\end{align}
where $\mu_q$ is the quark chemical potential. As discussed in Sec.~\ref{III}, $\delta a^2$ and $\delta \lambda$ represent the counterterms that ensure that the one-loop vacuum corrections do not shift the position of the minimum or the  vaccuum mass of the sigma. These counterterms are given by
\begin{align}
\delta a^2 &= -a^2\frac{(8g^4-12 \lambda^2-3\lambda^2 \ln[2])}{32 \pi \lambda}, \nn
\delta \lambda &= \frac{(16+8 \ln[g^2/\lambda])g^4-(18 +9  \ln[2])\lambda^2}{64 \pi^2}.
\end{align}
The self-energy at finite temperature and quark chemical potential, $\Pi(T_0,\mu_q)$, includes the contribution from both bosons  and fermions. In the high temperature approximation, it is given by~\cite{RMF}
\begin{eqnarray}
    \Pi(T_0,\mu_q)&=&-N_fN_cg^2\frac{T_0^2}{\pi^2}
    \Big[\text{Li}_2(-e^{\mu_q/T_0})
    \nonumber \\
    & & \hspace{1cm} + \text{Li}_2(-e^{-\mu_q/T_0})\Big]   
    + \frac{\lambda T_0^2}{2}.
    \label{fullselfenergy}
\end{eqnarray}
To implement superstatistics corrections, we proceed along the lines described in Sec.~\ref{III}. First, we substitute Eq.~(\ref{finalHTpotential}) into Eq.~(\ref{partitionfunctionnormal}). The partition function is obtained from Eq.~(\ref{partfuncsuper}) and the effective potential, including superstatistics effects is obtained from the logarithm of this partition function,
\begin{equation}
 V_{\text{sup}}^{\text{eff}}=-\frac{1}{\texttt{V}\beta}\ln[Z].
 \label{effpotsupLSMq}
\end{equation}
where the expression for $Z$ is now based on the effective potential Eq.~\eqref{finalHTpotential}. As a consequence, the effective potential of Eq.~(\ref{effpotsupLSMq}) has four free parameters. Three of them come from the original model, namely, $\lambda$, $g$ and $a$. The remaining one corresponds to the superstatistics correction, $N$. In the absence of superstatistics, the effective potential in Eq.~(\ref{finalHTpotential}) allows for second- as well as for first-order phase transitions, depending on the values of $\lambda$, $g$ and $a$ as well as of $T_0$ and $\mu_q$. For given values of $\lambda$, $g$ and $a$, we now proceed to analyze the phase structure that emerges when varying $N$, paying particular attention to the displacement of the CEP location in the $T_0$, $\mu_q$ plane. 

\begin{figure}[t]
 \includegraphics[scale=0.52]{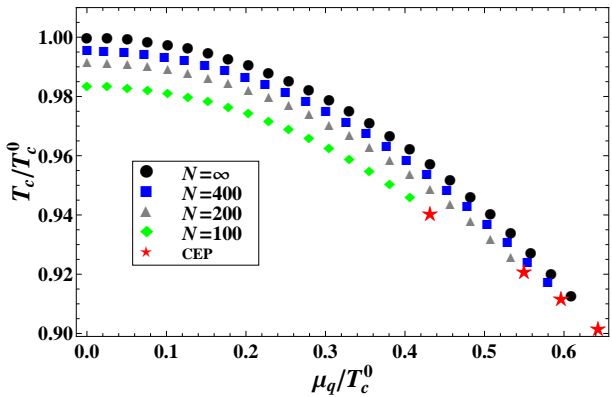}
 \caption{Effective QCD phase diagram calculated with $a=133$ MeV, $g=0.51$ and $\lambda=0.36$, for different values of $N$. The star shows the position of the CEP which moves towards larger values of $T$ and lower values of $\mu_q$, as $N$ decreases.}
 \label{fig3}
\end{figure}

Figure~\ref{fig3} shows the effective QCD phase diagram calculated with $a=133$ MeV, $g=0.51$ and $\lambda=0.36$ for different values of the number of subsystems making up the whole system, $N$. For the different curves, the star shows the position of the CEP. Notice that this position moves to larger values of $T$ and lower values of $\mu_q$, with respect to the CEP position for $N=\infty$, that is, without superstatistics effects, as $N$ decreases.

\begin{figure}[t!]
 \includegraphics[scale=0.52]{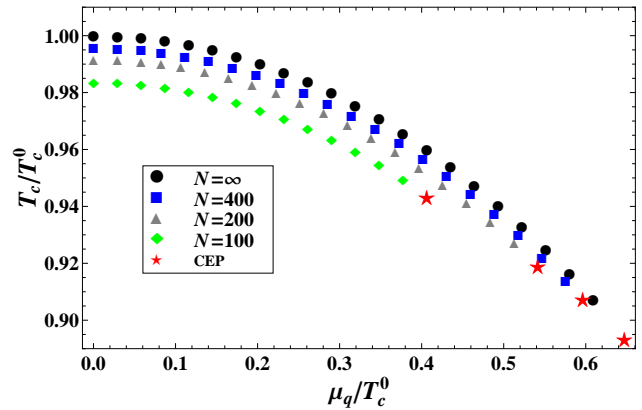}
 \caption{Effective QCD phase diagram calculated with $a=133$ MeV, $g=0.63$ and $\lambda=0.4$, for different values of $N$. The star shows the position of the CEP which moves towards larger values of $T$ and lower values of $\mu_q$, as $N$ decreases.}
 \label{fig4}
\end{figure}

Figure~\ref{fig4} shows another example of the effective QCD phase diagram, this time calculated with $a=133$ MeV, $g=0.63$ and $\lambda=0.4$. Notice that the systematics of the CEP displacement for this case are the same as for the case described in Fig.~\ref{fig3}. 

\section{Summary and outlook}\label{concl}

In this work, we have studied the effect of superstatistics for chiral symmetry restoration and in particular for the position of the CEP in an effective description of the QCD phase diagram. 

We have implemented the superstatistics scenario to account for fluctuations in temperature in a system that initially can be considered as not fully thermalized and made up of a given number of subsystems $N$. We have not considered fluctuations in the chemical potential, which have been included to study the CEP position in the Nambu--Jona-Lasinio model in Ref.~\cite{Wilknew} We chose to describe these fluctuations in terms of a $\chi^2$ distribution for the inverse temperature. 

The analysis is based on the superstatistics modification to the system's partition function to first order in $1/N$. To study these effects, we first resorted to finding the change in the critical temperature for $Z_2$ symmetry restoration in a theory with a self-interacting real boson field. We found that this critical temperature decreases as $N$ decreases. We then studied the LSMq at finite temperature and quark chemical potential. We found that the pattern for chiral symmetry restoration at high temperature and density changes from the case where $N=\infty$ (completely thermalized volume) to the case where the system is made up of a finite number of subsystems. In particular, the CEP position  moves toward larger values of $T$ and lower values of $\mu_q$ as $N$ decreases. For the analyses, in both cases, the free energy is computed beyond the mean-field approximation, including the plasma screening effects. All matter corrections are made in the high temperature approximation. 

Notice that one can wonder what is the proper way to average over the possible different subsystems' temperatures. In principle there are two alternatives: (a) computing the partition function as the trace of a modified Boltzmann factor coming from first averaging the possible temperature values or (b) computing first the trace of the Boltzmann factor for each subsystem and then averaging over the different subsystem’s temperatures. These two options were discussed for example in Ref.~\cite{Beck2004-I} and were called superstatistics type-A and -B, respectively. In type-A superstatistics one works with un-normalized Boltzmann factors $e^{-\beta E}$ that are averaged over $\beta$ with the distribution $f(\beta)$ and the normalization is carried out at the end by performing the integration over $E$. In type-B superstatistics one works with locally normalized distributions $p(E)=1/(Z(\beta))e^{-\beta E}$ to finally average over all $\beta$ with the distribution $f(\beta)$. Since in general the normalization constant $Z$ depends on $\beta$, the results will differ. However, case B can be easily reduced to case A by replacing the distribution $f(\beta)$ by a new distribution $\tilde{f}(\beta)= CZ^{-1}(\beta)f(\beta)$, where $C$ is a suitable normalization constant. In other words, type-B superstatistics with $f$ is equivalent to type-A superstatistics with $\tilde{f}$. In case one is not interested in the relation between the superstatistics
types, and since the normalization factor in the second case depends
on $\beta$, one can expect a different result when working with the same $f(E)$. Nevertheless, as explicitly worked out in Ref.~\cite{Obregon1}, different
$f(E)$'s lead to similar entropic factors when expanded to first order in $1/N$.

Also, notice that the first-order phase transitions start appearing when fermions become more relevant than bosons. This is bound to happen for a large enough baryon chemical potential. Therefore, fluctuations imply that some of the subsystems already reached a temperature above the critical temperature for this phase transition with the order parameter being zero while some others have a lower temperature with a nonzero order parameter value, whatever this may be. However, these transitions stay sharp and the question is how fluctuations influence the values for the baryon chemical potential and the temperature for fermions to become relevant. Our findings show that fermions become more relevant for lower values of the baryon chemical potential than they do for the case of the homogeneous system. To picture this result, as above, let $(\mu^0_c,T^0_c)$ and $(\mu_c,T_c)$ be the critical values for the baryon chemical potential and temperature for the onset of first-order phase transitions for the homogeneous and fluctuating system, respectively.  The parameter that determines when fermions become relevant is the combination $\mu^0_c/T^0_c$. Since our calculation for a single boson d.o.f. shows that the critical temperature decreases with a decreasing number of subsystems, this means that for the boson-fermion fluctuating system, fermions become relevant for $\mu_c/T_c \simeq \mu^0_c/T^0_c$, and thus for $\mu_c < \mu^0_c$.

To apply these considerations in the context of relativistic heavy-ion collisions, we recall that temperature fluctuations are related to the system's heat capacity by
\begin{equation}
 \frac{(1-\xi)}{C_v}=\frac{\langle (T-T_0)^2\rangle}{T_0^2},
 \label{heatcapacity}
\end{equation}    
where the factor $(1-\xi)$ accounts for deviations~\cite{Wilk:2009nn} from the Gaussian~\cite{landau} distribution for the random variable $T$. The right-hand side of Eq.~(\ref{heatcapacity}) can be written in terms of fluctuations in $\beta$ as
\begin{eqnarray}
 \frac{\langle (T-T_0)^2\rangle}{T_0^2}
 &=&
\frac{\beta_0^2-\langle\beta^2\rangle}{\langle\beta^2\rangle}\nonumber\\
 &=&\frac{\Big( \frac{\beta_0^2}{\langle\beta^2\rangle} \Big)^2\langle\beta^2\rangle-\beta_0^2}
{\beta_0^2}.
\end{eqnarray}
Notice that according to Eq.~(\ref{variance})
\begin{eqnarray}
\Big( \frac{\beta_0^2}{\langle\beta^2\rangle} \Big)^2&=&\left(\frac{1}{1+2/N}\right)^2\nonumber\\
&\simeq&1-4/N.
\label{notice}
\end{eqnarray}
Therefore, for $N$ finite but large 
\begin{equation}
 \frac{\langle(T-T_0)^2\rangle}{T_0^2}\simeq\frac{\langle\beta^2\rangle-\beta_0^2}{\beta_0^2},
 \label{T-beta}
\end{equation}
and using Eqs.~(\ref{variance}) and~(\ref{T-beta}), we obtain
\begin{equation}
\frac{\langle(T-T_0)^2\rangle}{T_0^2}=\frac{2}{N}.
\end{equation}
This means that the heat capacity is related to the number of subsystems by
\begin{equation}
 \frac{(1-\xi)}{C_v}=\frac{2}{N}.
 \label{C-N}
\end{equation}

To introduce the specific heat $c_v$ for a relativistic heavy-ion collision, it is natural to divide $C_v$ by the number of participants $N_p$ in the reaction. Therefore, Eq.~(\ref{C-N}) can be written as
\begin{equation}
 \frac{2}{N}=\frac{(1-\xi)}{N_p c_v}.
 \label{N-Np}
\end{equation}
In Ref.~\cite{Wilk:2009nn}, $\xi$ is estimated as $\xi=N_p/A$, where $A$ is the smallest mass number of the colliding nuclei.
Equation~(\ref{N-Np}) provides the link between the number of subsystems in a general superstatistics framework and a relativistic heavy-ion collision. It has been shown~\cite{Basu} that, at least for Gaussian fluctuations, $c_v$ is a function of the collision energy. Therefore, in order to make a thorough exploration of the phase diagram as the collision energy changes, we need to account for this dependence as well as to work with values of the model parameters $\lambda$, $g$ and $a$, appropriate to the description of the QCD phase transition. Work along these lines is currently underway and will be reported elsewhere.

\section*{Acknowledgements}

M. H., M. L. and R. Z. would like to thank Instituto de Ciencias Nucleares, Universidad Nacional Aut\'onoma de M\'exico for their warm hospitality during a visit in June-July 2018. A. A. and L. A. H. would like to thank the Physics Derpartment, PUC and CIDCA for their warm hospitality during a visit in July 2018. This work was supported by Consejo Nacional de Ciencia y Tecnolog\'ia grant number 256494, by Fondecyt (Chile) grant numbers 1170107, 1150471, 11508427, Conicyt/PIA/Basal (Chile) grant number FB0821. R. Z. would like to acknowledge support from CONICYT FONDECYT Iniciaci\'on under grant number 11160234.

\end{document}